# Field Experimental "Star Type" Metropolitan Quantum Key Distribution Network

Wei Chen, Zheng-F Han, Tao Zhang, Hao Wen, Zhen-Q Yin, Fang-X Xu, Qing-L Wu, Yun-Liu,
Yang Zhang, Xiao-F Mo, You-Z Gui, Guo Wei, Guang-C Guo.

*Abstract*—Quantum key distribution (QKD) network has recently attracted growing attentions. Due to the special characteristics of quantum information, to build a full-connectivity QKD network without trusted relays is a stimulating challenge. In this letter, we report on the first realization of QKD network without trusted relays which covers metropolis in the commercial backbone optical fiber networks. The star topology four-user QKD network automatically addresses the quantum signal with a quantum router (QR) and every user in the network can receive and distribute quantum keys to any others simultaneously. The longest and the shortest length of fibers between two geographically separated nodes are 42.6km and 32km respectively, and the maximum average quantum bit error rate (QBER) is below 8%. This result opens a new possibility for the use of QKD into existing network.

## I. INTRODUCTION

Quantum key distribution (QKD) is a promising candidate technology for the next generation security solution, which exploits the fundamental physical laws of nature to allow the geographical separated couples to share absolutely secure keys [1]. There have been many point-to-point realizations of QKD since its first demonstration in 1989, and some of them have successfully operated in telecom dark fibers [2]-[4]. Recently more and more researchers have aimed at developing an applicable global QKD network for unconditional secure key distribution.

In the past few years, several QKD network topologies have been proposed: the passive splitter network [5], the optical ring network [6], the optical-switch active routed network [7], and wavelength-addressed bus or tree network [8], [9]. Some limitations can be found in these topologies, such as the user in most of these nets can not exchange keys with all of others at will, and the performance of single user is influenced by others in the network or will decline sharply as user number increasing. Besides those schemes the multi-particles entanglement QKD network is another candidate [10], however it is still far away from practical applications due to difficult multi-particles entanglement source.

In this letter, we take star topology QKD network topology based on wavelength-division multiplexing (WDM) [11] to implement a four-user QKD net was built in the commercial backbone telecom fiber network of China Netcom Company limited (CNC) in Beijing. Network addressing is automatically processed by the "Quantum Router" (QR) in the center of the net. In this network, one user can directly exchange the quantum keys to any other one without trusted relays. The performance of every user is almost equivalent and will not deteriorate as the network expanding.

## II. TOPOLOGY OF MULTI-USER QKD NETWORK

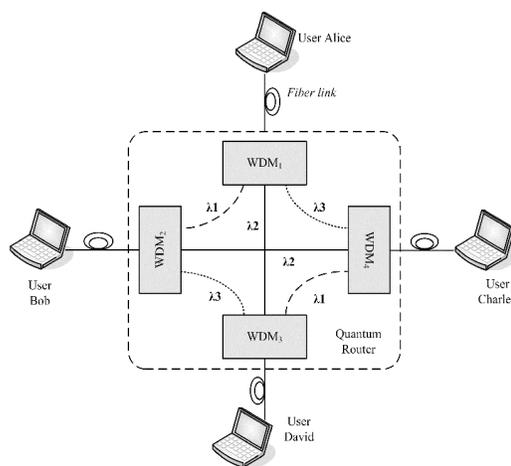

Fig. 1. Structure of four-user quantum router and the star network topology.

The net uses star topology with a QR in the center. As an example, we use four-user QKD network to describe its function details. The structure of four-port quantum router which is composed of three wavelength WDMs as shown in Fig. 1. When Alice wants to perform quantum communication with Bob, she

Manuscript received July 27, 2007. This work was supported in part by National fundamental Research Program of China under Grant No 2006CB921900, the National Science Foundation of China under Grant No. 60537020 and 60621064, and the Knowledge Innovation Project of Chinese Academy of Sciences & Chinese Academy of Sciences and International Partnership Project.

W. Chen, Z.F. Han, T. Zhang, H. Wen, Z.Q. Yin, F.X. Xu, Q.L. Wu, Y. Liu, Y. Zhang, X.F. Mo, Y.Z. Gui, and Guang-C Guo. are with the Key Lab of Quantum Information, CAS, USTC, Hefei, China (email: zfhan@ustc.edu.cn ).
W. Chen, H. Wen, Q. L. Wu, and G. Wei are with the Department of Electronic Engineering and Information Science, USTC, Hefei ,China.

sends out the photons of wavelength $\lambda_1$, the light will be separated by $WDM_1$ and transmitted to $WDM_2$, then be assembled to the link to Bob by $WDM_2$. When Alice wants to exchange the quantum keys with Bob, David and Charley simultaneously, she can send out the photons of $\lambda_1$, $\lambda_2$ and $\lambda_3$ at the same time. The communication procedure of other users is the same as Alice. There are only two WDMs added into the point to point link of Alice to Bob and the system security is not deduced, so that arbitrary point-to-point QKD protocols can be used by Alice and Bob to perform QKD.

The network can be extended to N users by using an N-port QR. The connection scheme concurs with the edge coloring theorem in graph theory [11]. Each WDM, link, and wavelength corresponds to a vertex, edge and color in the graph respectively. Duet to the edge coloring theorem, it needs N-1 colors to render a complete graph with N vertexes when N is even, and N colors when N is odd. This means for an N-port QR, each WDM should be an N-1(when N is even) or N (when N is odd) wavelength multiplexer. It is worth mentioning that according to this structure, no matter how many users are added, the photon can be sent from one node to any other node only passing through two WDMs and the network performance will not degrade theoretically. When multi couples are working simultaneously, the crosstalk is primarily limited by the isolation of WDMs. For commercial WDM products N can exceed 40 and the isolation for adjacent (non-adjacent) channel can over 30dB (45dB). For a QKD network covering a region of 50km diameter, the ratio of the crosstalk to effective signal is less than 0.06%. This indicates that, in such a QKD network, the error due to crosstalk is negligible comparing with other factors such as dark counts of single photon detector (SPD) or imperfect interference.

III. EXPERIMENT SETUP AND RESULT

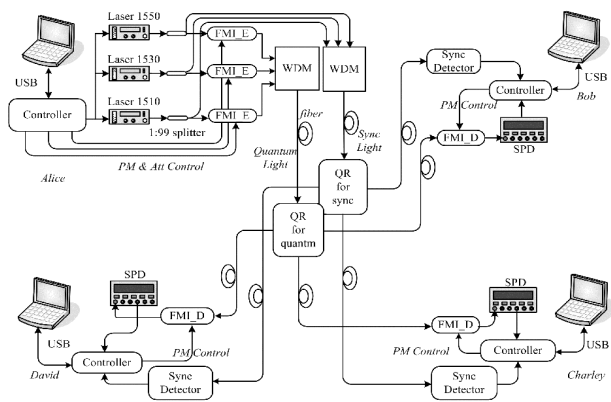

Fig. 2. System diagram of four-user QKD network. FMI_E: F-M interferometer encoder; FMI_D: F-M interferometer decoder; Controller: the electronic control board. PM Control: phase modulation driving voltage; Att Control: electronic variable optical attenuator control signal; QR: quantum router; SPD: single photon detector.

Fig. 2 shows the experiment setup of the four-user network in which three wavelengths are necessary. Alice acts as sever who sends quantum keys to client Bob, David and Charley. Alice uses the attenuated light from three DFB lasers of 1550nm, 1530nm and 1510nm (Advanced Photonic System) as photon sources. Each receiver is assigned a wavelength as its unique address. The light from laser is divided into quantum signal and sync signal for triggering the detection system. There are two 4-port QRs of the same structure in the center of star topology, one for the quantum light and the other for the sync light. The three-wavelength quantum lights from Alice are multiplexed into a fiber link and the three sync lights are multiplexed into the other fiber. Each receiver of Bob, David and Charley also has two fiber links connecting to the QR for quantum and sync light. The classical communication is directly accomplished by the fiber optic transceiver with individual fibers.

Faraday-Michelson (F-M) QKD systems [4] is used in the network The system can automatically compensate for the birefringence of fiber so that it can work stably for a long time without active polarization recovery. Strict BB84 protocol [12] is used to perform QKD in the experiment. Alice applies the driving signals to its phase modulator ($PM_A$) to randomly encode the information unto the photon phases of 0, $\pi/2$, $\pi$ and $3\pi/2$. Then the second light pulse output from the interferometer is attenuated to average 0.1photon/pulse and sent into the quantum channel. The receiver also randomly selects the four phases of $PM_B$, and the interference result is delivered into an InGaAs/InP avalanche single-photon detector (id Quantique id200). According to BB84 protocol, the basis was divided into two sets of {0, $\pi$} and {$\pi/2$, $3\pi/2$} in which phases 0 and $\pi/2$ denotes key 0, $\pi$ and $3\pi/2$ denotes key 1. The four-phase random selection and single-SPD scheme used in the receiver immunizes our system from fake-state attack [13], and its security is better than two-phase selection and double-SPD setup. The couples declare the basis set they selected in every time bin when the receiver detected photons and the sifted key of the same basis set are kept.

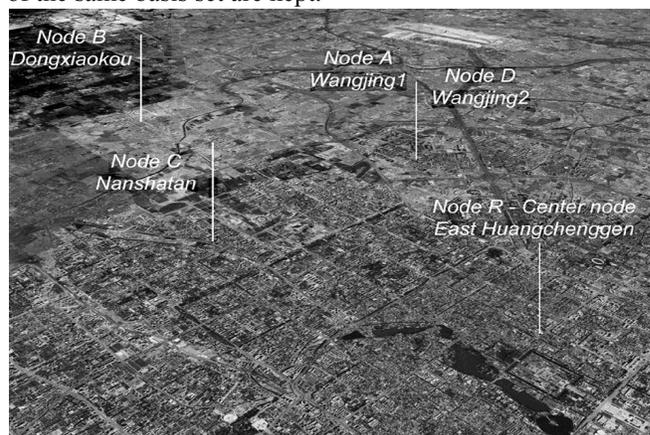

Fig. 3 Node positions of QKD network in Beijing. The quantum router is located in the center node (Node R). The area below center node is the Forbidden City and the Tiananmen Square. The fiber length of A to R is 19.88km, B to R is 22.8km, C to R is 12km.

Fig. 3 shows the actual positions of the four users accessed into the backbone fiber network of CNC, Beijing branch. The transmitter Alice (node A) and receiver Bob (node D) are in Wanjing bureau; receiver David (node B), Charley (node C) is in the Dongxiaokou and the Nashantan bureau respectively. The QR is placed in the East Huangchenggen bureau. In order to validate the connectivity of the network, we also perform the QKD session from node D to node C. Table I lists the fiber line parameters and the performance of the QKD network. The QBER and bit rate are average results of the long time continuous operation. We tested the single link mode (only single wavelength QKD session running) and the concentration mode (the three wavelength QKD sessions running simultaneously). The maximum QBER of the four-user QKD network is below the basic security threshold of 11.5% [14]. When taking imperfect realizations of BB84 and special attacks such as photon number splitting (PNS) [15] into account, the security QKD sessions will have more restrictions. In this situation, refined QKD protocols, for example, decoy state [16] can be used to guarantee the security. In order to demonstrate it, a two-intensity decoy QKD session in the link A-R-C has been accomplished. In this experiment, the average photon number of the decoy state and signal state are $\mu=0.2$ and $v=0.6$ per pulse respectively. The repetition rate of the laser is 1MHz and the final security key rate is $6.784 \times 10^{-6}$ per pulse. Due to the independence of the end-to-end QKD realizations and the network topology, here we only list the QBER as a reference. The decoy state experiment will be described in another article.

TABLE I
WAVELENGTH ARRANGEMENT AND THE PERFORMANCE OF THE NETWORK. A-R-B is the link from the node A (Alice) pass through the node R (center node with the QR) and then to node B (David) and others are similar. QBER1 and sifted key rate1 (QBER2 and sifted key rate 2) is the average result of single mode (concentrate mode).

| Result | A-R-B | A-R-C | A-R-D | D-R-C |
|---|---|---|---|---|
| Fiber Length (km) | 42.68 | 31.88 | 39.76 | 31.88 |
| Wavelength (nm) | 1530 | 1510 | 1550 | 1530 |
| Path Loss (db) | 16.44 | 11.63 | 15.59 | 10.68 |
| Fringe Visibility | 99.77% | 97.44% | 99.75% | 99.98% |
| SPD Dark Count | $9.7\times10^{-6}$ | $5.2\times10^{-6}$ | $1.2\times10^{-5}$ | $5.2\times10^{-6}$ |
| $QBER_1$ | 7.7 % | 4.1 % | 6.6 % | 2.4% |
| Sifted key rate$_1$ (bps) | 33 | 57 | 35 | 48 |
| $QBER_2$ | 7.5% | 4.4% | 6.6% | / |
| Sifted key rate$_2$ (bps) | 32 | 53 | 36 | / |

TABLE II
ISOLATION OF QUANTUM ROUTER.

| Alice Input(nm) | 1550 output(dB) | 1530 output(dB) | 1510 output(dB) |
|---|---|---|---|
| 1550 | -2.45 | -38.66 | -43.35 |
| 1530 | -44.59 | -2.27 | -43.75 |
| 1510 | -51.01 | -44.80 | -1.76 |

The isolation and the insertion loss of the QR for the quantum signal are listed in Table II. The precision of the isolations is limited by our detection apparatus. According to the white paper of WDM, the adjacent and non-adjacent channel isolation of single WDM is more than 30dB and 45dB respectively so that the isolation of the QR should be more than 60dB. Comparing the performance of the single mode with the concentration mode in table 1, we can see that there are little differences between two modes. This result indicates that the crosstalk of this QKD network is little enough in the experiment, and is consistent with the theoretical analysis.

## IV. CONCLUSION

This experiment result indicates the feasible network topology is appropriate for QKD and feasible to be integrated into nowadays fiber networks. Beside that, meeting basic requirements of quantum communication, the infrastructure can be extensively used in future's quantum network. However, to integrate the classical and quantum signals to a single fiber link and to deduce the cost of network are the important projects in the following research.

ACKNOWLEDGMENT

The authors thank the CNC for supplying the fields and fibers in Beijing.